\documentclass[proceedings]{JHEP}
\usepackage{epsfig}
\conference{Heavy Flavours 8, Southampton, UK, 1999}
\title{The measurement of sin(2$\beta$)}
\author{D. Bortoletto \\
Purdue University, West Lafayette IN 47907-1396, USA \\
\email{daniela@physics.purdue.edu}}

\abstract{
Since the first observation in 1964, CP violation remains one of the
most elusive aspects of the standard model. The CDF collaboration
has reported the first evidence of CP violation in the B system
using the world's largest sample of $B \rightarrow J/\psi K^0_S $ decays.
The direct measurement of sin$(2\beta)$=0.79$^{+0.41}_{-0.44}$ (combined
statistical
and systematic error)
agrees with the standard model predictions.
New data collected from the B-factories and from the upgraded experiments
at the Tevatron should allow a more precise measurement
of $\sin 2\beta$ in the near future.}

\begin{document}

\section{Introduction}

The violation of invariance under charge-conjugation and parity (CP)
transformations was first observed in the neutral Kaon system in 1964
\cite{obs1964}. In the standard model with three generations of fermions,
CP violation arises from a single
complex phase in the mixing matrix for quarks. The CKM
matrix, $V_{CKM}$, is a unitary matrix \cite{ckm} that transforms the mass
eigenstates to the weak eigenstates:
\begin{eqnarray*}
\lefteqn
{V_{CKM}=\left(\begin{array}{ccc}
V_{ud} & V_{us} & V_{ub} \\
V_{cd} & V_{cs} & V_{cb} \\
V_{td} & V_{ts} & V_{tb} \\
\end{array}\right)
 \approx  }  \\
& & \left(\begin{array}{ccc}
1-\frac{\lambda^2}{2} & \lambda & A\lambda^3(\rho-i\eta ) \\
-\lambda & 1-\frac{\lambda^2}{2} & A\lambda^2 \\
 A\lambda^3(1-\rho-i\eta ) & -A\lambda^2  &1 \\
\end{array}\right) + O(\lambda^4)
\end{eqnarray*}
It is useful to express the CKM matrix in terms of the four
Wolfenstein \cite{wolf} parameters: $\lambda$, $A$, $\rho$ and $\eta$.
The sine of the Cabibbo angle, $\lambda$, is measured in semileptonic
kaon decays, $\lambda=|V_{us}|=0.2196\pm 0.0023$, and plays the role
of an expansion parameter. $A$ can be determined in
$b\rightarrow c$ decays since
$A=V_{cb}/\lambda^2$. Only $\lambda$ and $A$ are measured
precisely while $\eta$ and $\rho$
must still be determined. The parameter $\eta$ represents
the CP-violating phase and must be
different from zero
to accommodate CP violation in the standard model.

The condition of unitarity between the first and third columns:
$$
V_{ud}V^*_{ub}+V_{cd}V^*_{cb}+V_{td}V^*_{tb}=0
$$
can be represented geometrically as a
triangle in the $\rho$-$\eta$ plane. The apex of the triangle
has coordinates ($\rho$, $\eta$) as shown in figure 1.
The three angles of the unitarity
unitarity triangle are :
\begin{eqnarray*}
\alpha & = & \arg \left[- \frac {V_{td}V^*_{tb}}{ V_{ud}V^*_{ub}}
\right]  \\
\beta & = & \arg \left[- \frac {V_{cd}V^*_{cb}}{ V_{td}V^*_{tb}}
\right]  \\
\gamma & = &\arg \left[- \frac {V_{ud}V^*_{ub}}{ V_{cd}V^*_{cb}}
\hfill \right]
\end{eqnarray*}
Measurements
of weak decays of $B$ hadrons determine the magnitudes of the three sides
while CP asymmetries in B meson decays determine the three angles.
\EPSFIGURE[ht]{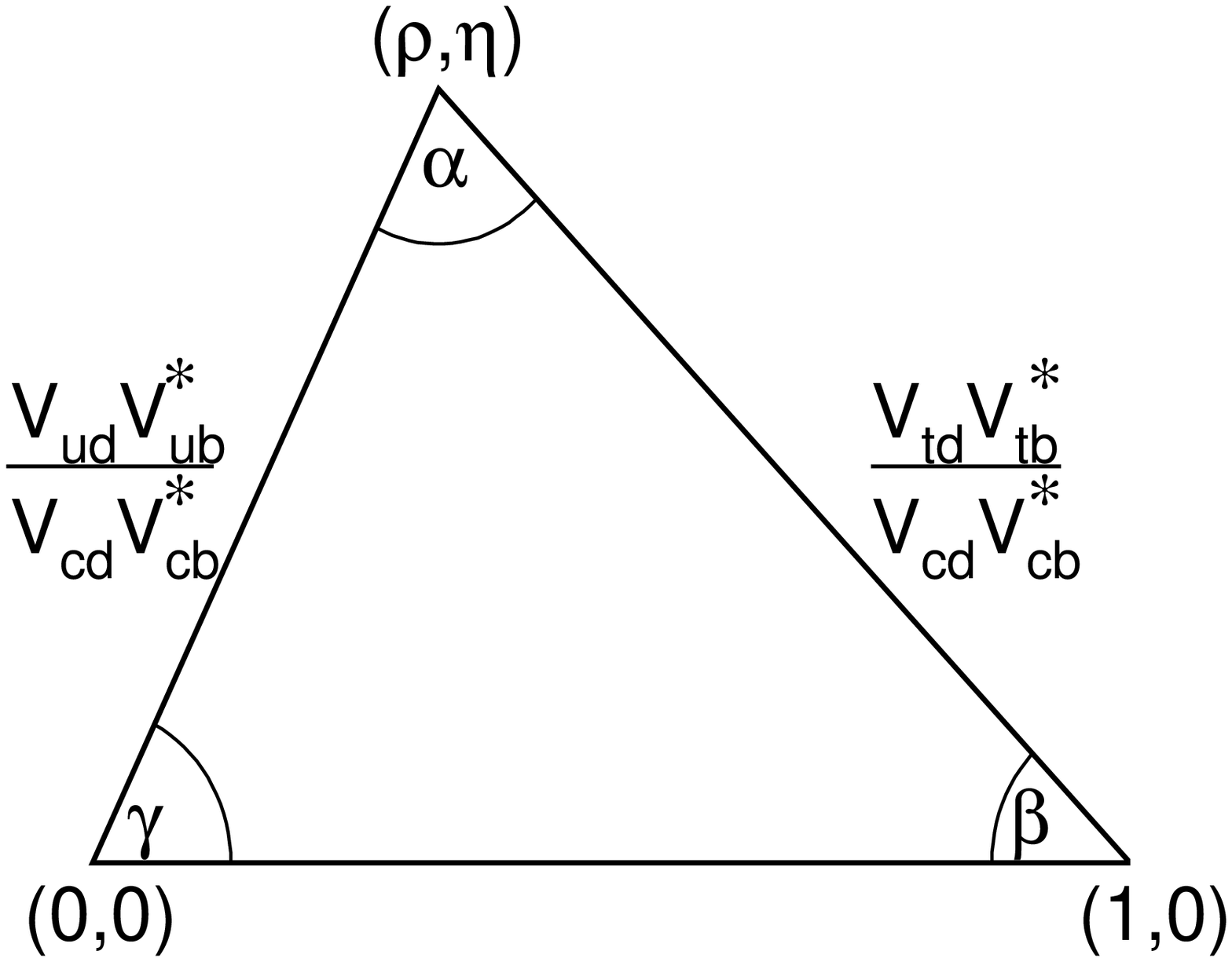,width=8cm}
{The unitary triangle in the complex $\rho-\eta$ plane.}
The goal of B physics in the next decade is to measures
both the sides and angles of the unitarity triangle and test consistency
within the standard model.

\section{CP violation in the neutral B meson system}

The neutral b meson system, $B^0(\bar b d)$, is not an eigenstate of
CP. The standard phase space convention defines:
\begin{eqnarray*}
CP|B^0> & = & -|\bar B^0> \\
CP|\bar B^0> & = & -|B^0>
\end{eqnarray*}
Similarly to the neutral $K$ mesons,
the mass eigenstates of the neutral B system are not the flavor
eigenstates. The mass eigenstates are
called heavy and light and they are
denoted by $B_H$ and $B_L$:
\begin{eqnarray*}
|B_L> & = & p|B^0>+q|\bar B^0> \\
|B_H> & = & p|B^0>-q|\bar B^0>
\end{eqnarray*}
with $\Delta m_d=M_H-M_L$ and $\Delta \Gamma =\Gamma_H-\Gamma_L$
the mass and width difference respectively. While
the two neutral K mesons have very different lifetimes,
the lifetime difference in the neutral B system is
negligible. Therefore the flavour of the B meson at production
must be determined by tagging.

The possible manifestation of CP violation can be classified according
to three main categories: CP violation in the decays,
CP violation in mixing and CP violation in the
interference between decays with and without mixing.

CP violation in the decays can occur both in charged and
neutral decays when the amplitude for a decay and its CP conjugate
mode have different magnitude. This is often called direct CP violation
and can be studied by comparing the decay rates $\Gamma (B\rightarrow f)$
with $\Gamma(\bar B\rightarrow \bar f)$, where $f$ and $\bar f$ are
CP-conjugate final states. If $A$ and $\bar A$ are the amplitudes for
the process $B \rightarrow f$ and $\bar B \rightarrow \bar f$, we have
CP violation in the decay if
$$
\left | \frac{A}{\bar A} \right | \neq 1.
$$
If this condition is met the CP asymmetry
$A_{CP}$ can be defined as :
$$
A_{CP}=\frac{\Gamma(\bar B\rightarrow \bar f)-\Gamma(B \rightarrow  f )}
{\Gamma(\bar B\rightarrow \bar f) + \Gamma(B \rightarrow  f )}
$$
Since in the standard model the amplitude for a decay
and its CP conjugate differ at most by a phase factor, we can observe
CP violating effects
only if there are at least two interfering amplitudes. In the case of the decay
$\bar B^0 \rightarrow K^- \pi^+$ such interference can be provided by
the tree and the penguin contributions.

CP violation in mixing can occur when two neutral mass eigenstates
cannot be chosen to be CP eigenstates. This is also called indirect
CP violation
and it is measured by the parameter $\epsilon$ in kaon decays.
Mixing provides an interfering amplitude that can potentially produce
CP violation. For example, CP violation in the kaon system is
largely due to mixing. The condition for CP violation in mixing is that
$$
\left | \frac{q}{p} \right |  \neq 1
$$
In the neutral B system, indirect CP violation is expected to be
small, of ${\cal O}(10^{-4})$.

CP violation in the interference between decays with and without mixing
can occur when there are CP final states, $f_{CP}$, that are
common to $B^0$ and $\bar {B^0}$. In these cases the process
$B^0 \rightarrow \bar B^0 \rightarrow f_{CP}$
can interfere with the direct decay $B^0 \rightarrow f_{CP}$.
Let us assume that the final state $f_{CP}$ is an eigenstate of CP with
eigenvalues $\eta_{CP}$= $\pm $1.
CP violation produces a non exponential decay probability:
\begin{eqnarray*}
\lefteqn
{\left | A(B^0(t) \rightarrow f_{CP}) \right |^2  \approx  } \\
& & e^{-\Gamma t} [ 1- \eta_{CP}(f)\sin(\Delta m_d t) \sin2(\phi_M-\phi_D)]
\end{eqnarray*}
\begin{eqnarray*}
\lefteqn
{\left | A(\bar B^0(t) \rightarrow f_{CP}) \right |^2  \approx } \\
& & e^{-\Gamma t} [ 1+ \eta_{CP}(f)\sin(\Delta m_d t) \sin2(\phi_M-\phi_D)]
\end{eqnarray*}
where $\phi_M$ and $\phi_D$ are the mixing phase and the direct weak
phase respectively.

The CP asymmetry as a function of time is defined as:
\begin{eqnarray*}
A_{CP}(t) & = & \frac{\frac{d\Gamma}{dt}(\bar B^0 \rightarrow f_{CP})-
\frac{d\Gamma}{dt}(B^0 \rightarrow f_{CP})}
{\frac{d\Gamma}{dt}(\bar B^0 \rightarrow f_{CP})
+\frac{d\Gamma}{dt}(B^0 \rightarrow f_{CP})}  \\
& = & \eta_{CP}(f)\sin2(\phi_M-\phi_D)\sin\Delta m_d t
\end{eqnarray*}

We can also compute the time integrated asymmetry as:
$$
A_{CP}=\eta_{CP}(f)\frac{x_d}{1+x_d^2}\sin2(\phi_M-\phi_D)
$$
where $x_d=\Delta {m_d} / \tau(B^0)=0.732 \pm 0.032$ \cite{PDG} is
the $B^0-\bar B^0$ mixing parameter.
Such a time integrated asymmetry can be measured in a hadron collider where
the B mesons are produced uncorrelated. At a B-factory operating
at the $\Upsilon(4S)$ the B mesons are produced
as a $B^0$ and a $\bar B^0$ in a correlated C=$ -$ 1 quantum state.
Therefore, the time
integrated CP asymmetry vanishes since the two B mesons evolve coherently;
until
one of them decays, we always have a $B^0$ and a $\bar B^0$.
At a B-factory operating at the
$\Upsilon(4S)$, it is crucial to measure the CP asymmetry as a function of
the time difference, $\delta t=t_{CP}-t_{tag} $, between the B decaying
into a CP eigenstate and the tagged B final state.
At the Tevatron the B mesons are produced
incoherently and the asymmetry can be measured either as a time dependent
or a time integrated quantity. The time dependent analysis is more powerful.
First, there is more statistical power in the time dependent asymmetry;
decays at low lifetime exhibit a small asymmetry because there has not
been
enough time for mixing to generate the interference which leads to CP
violation. In fact $A_{CP}(t)$ has a maximum at
$t \approx 2.2$ lifetimes. Moreover, a substantial fraction of the
combinatorial background occurs at short proper time.

\section{Determination of $\sin 2\beta$ }

The angle $\beta$ can be determined by studying different
types of b decay modes.
The best determination of $\sin 2\beta$ can be achieved by studying color
suppressed decays such as $b \rightarrow c \bar c s$.
The golden mode is the decay $B^0/\bar B^0 \rightarrow J/\psi K^0_s$.
The dominant penguin contribution has the same weak phase as the tree
amplitude\cite {bluebook}. The only term with a different weak phase
is suppressed by ${\cal O}(\lambda^2)$. Therefore the extraction of
$\beta$ from the measurement of the asymmetry in
$B^0/\bar B^0 \rightarrow J/\psi K^0_s$ suffers negligible
theoretical uncertainties.

Cabibbo suppressed modes such as $b \rightarrow c \bar c d$
can also be used to determine
$\sin 2\beta$. The tree amplitude
for $B^0/\bar B^0 \rightarrow D \bar D$ is proportional to
$-\sin 2 \beta$. The contribution of penguin graphs with different
weak phases is potentially significant in these decay modes
since the tree contribution is Cabibbo suppressed.
A determination of the possible phase shift due to the
penguin contribution will be highly
model dependent and have large theoretical uncertainties
since it will be dominated by low energy hadronic effects.
Progress on measuring $\beta$ through these decays will be possible only by
comparing a variety of final states.

In penguin dominated modes such as $b\rightarrow s \bar s s $
or $b\rightarrow d \bar d s $ the tree contribution
is absent or is both Cabibbo and color suppressed. Predictions can
be obtained for a few modes such as $B\rightarrow \phi K^0_s$ but the
theoretical status is unclear.

At hadron colliders the golden mode
$B \rightarrow J/\psi K^0_s$ is especially interesting experimentally
because the $J/\psi \rightarrow \mu^+ \mu^-$ decay mode gives a
unique signature and allows for a powerful trigger.
Therefore the decay $B^0 \rightarrow J/\psi K^0_S$ can be
reconstructed with an excellent signal to background ratio. The
interference of the
direct decays $B^0 \rightarrow J/\psi K^0_S$ versus
$B^0 \rightarrow \bar B^0 \rightarrow J/\psi K^0_S$ gives rise to a CP
asymmetry that measures $\sin 2\beta$ :
$$
A_{CP}(t)=\frac{ \bar B^0(t) - B^0(t) }{ \bar B^0(t) + B^0(t) }
=\sin 2\beta \sin \Delta m_d t
$$
The time integrated asymmetry is:
$$
A_{CP}=\frac{x_d}{1+x_d^2}\sin2\beta
$$

To measure the asymmetry we have to identify the flavor of the
B meson at the time of production. The tagging algorithms are evaluated
in terms of efficiency $\epsilon$ for assigning a flavor tag and the
probability that the tag is correct.  The quality of the tag is evaluated
by defining the tagging dilution as $D=(N_R-N_W)/(N_R+N_W)$ where $N_R(N_W)$
is the number of right (wrong) tags. The observed asymmetry is reduced
with respect to the true asymmetry by the dilution $D$:
$A^{obs}_{CP}=DA_{CP}.$  Therefore, the maximum sensitivity can be achieved
when the dilution is large. A perfect tagging algorithm will have
a dilution $D=1$.
The statistical uncertainty on $\sin 2\beta$ is inversely proportional to
$\sqrt{\epsilon D^2}$ and to
$\sqrt{ S^2/(S+N_{bck})}$ where $S$ is the number of signal
events and $N_{bck}$ is the number of backgrounds events within three
standard deviations of the B mass. Therefore the data sample should be
chosen to maximize the signal and minimize the background.

Searches for a CP violating asymmetry using $J/\psi K^0_S$ samples have been performed
by the OPAL \cite{opal} and by the CDF \cite{oldcdf} collaborations which have reported
$\sin 2\beta=3.8^{+1.8}_{-2.0}\pm 0.5$  and 
$\sin 2\beta =1.8 \pm 1.1 \pm 0.3 $ respectively.
  
By combining indirect measurements it is possible to constrain the value
of $\sin 2\beta$. Currently several analysis find that the standard model
prefers large positive values of $\sin 2 \beta$. One recent global fit
finds $\sin 2 \beta= 0.75 \pm 0.09$ \cite{mele}.

\section{The CDF Detector}

The CDF detector is described in detail elsewhere~\cite{CDF}.
The measurement of the CP asymmetry requires reconstruction of the decay
mode $B^0 / \bar B^0 \rightarrow J/\psi K^0_S$ with good signal over
background, the measurement of the proper time
$t$, and the determination of the $B$ flavor at production.
The crucial detector features to achieve these requirements
include a four layer silicon
microstrip detector (SVX), a large volume drift chamber, and excellent
electron and muon identification. The SVX measures
the impact parameter, $d$, of charged tracks with a resolution that allows
the precise determination of the B meson proper decay time $t$.
The impact parameter resolution
for charged tracks is $\sigma_d= (13 + 40/P_T)$ $\mu$m, where
$P_T$ is the magnitute
of the component of the momentum of the track transverse to the beam line
in units of GeV/c.

The CTC has a radius of 1.4 m and is operating in a 1.4 T axial magnetic B
field. The CTC provides excellent track reconstruction efficiency and good momentum resolution,
$(\delta P_T/P_T)^2=(0.0066)^2 \oplus (0.0009)^2$,
 which allows $B$ meson reconstruction
with a mass resolution of $ \approx$
10 MeV/c$^2$. Electron ($e$) and ($\mu$) detectors in the central
rapidity region ($|y|<1$) allow B meson detection and provide
triggers using semileptonic $B$ decays or $B\rightarrow J/\psi X$ and
$J/\psi \rightarrow \mu^+ \mu^-$.

\section{The CDF Data Sample}

CDF exploits the large B cross section at the Tevatron to obtain a large
sample of $J/\psi K^0_S$ decays to measure $\sin 2 \beta$. The analysis
reported here
uses the entire Run I data sample of 110 pb$^{-1}$ and is described in
detail
in \cite{newcdf}.
The $J/\psi $ sample is identified by selecting two oppositely charged
muon candidates, each with $P_t>1.4$ GeV/c. A $J/\psi$ candidate is defined
as a $\mu^+ \mu^-$ pair within $\pm 5 \sigma$ of the world average mass
of 3.097 GeV/c$^2$, where $\sigma$ is the mass uncertainty for each event.
The $K^0_S$ candidates are found by matching oppositely charged tracks
assumed to be pions. The $K^0_S$ candidates
are required
to have $P_T(K^0_S) > 700$ MeV/c and
to travel a significant distance $L_{xy} > 5 \sigma$ from the
primary vertex, where $L_{xy}$ is the 2-dimensional flight distance
in the plane perpendicular to the direction of the beam.
The $\mu^+ \mu^-$ and $\pi^+ \pi^-$
are constrained to the appropriate masses and separate decay vertices.
A $K^0_S$ candidate is constrained to point back to the
$B^0$ meson decay point, and the $B^0$ meson candidate
is constrained to point back to the
primary vertex. In order to further improve the
signal-to-background ratio, B candidates are required to have
$P_T(B)$ above 4.5 GeV/c.

The data are divided into two samples which are called SVX and the
non-SVX samples.
The SVX sample requires both muon candidates to be well measured
by the silicon vertex detectors.
The two dimensional flight distance, measured in the plane perpendicular
to the beam, can be used to calculate the proper decay time $ct$, which
is the projection of the displacement along the B momentum.
The B candidates in this sample have a
precise proper decay time resolution of
$\sigma_{ct}\approx 60$ $\mu$m. The non-SVX sample contains events
in which one or both of the muon candidates are not measured in the
silicon vertex detector. B candidates in this sample have low proper 
decay time resolution of
$\sigma_{ct}\approx 300-900$ $\mu$m.
The lifetime information for both the SVX and the non-SVX B candidates
is used to evaluate the asymmetry as a function
of proper time in order to reduce the statistical
uncertainty in $\sin 2\beta$.

The $J/\psi K^0_S$ mass from the vertex and mass-constrained
fit is used to define the normalized
mass $M_N=(m_{\mu\mu\pi\pi}-M_0)/\sigma_{fit}$.
The uncertainty in the fit, $\sigma_{fit}$,
is typically $\approx$ 10 MeV/c$^2$, and the world average
neutral $B^0$ mass, $M_0$ is  5.2792 GeV/c$^2$.
The normalized mass distributions for the SVX and non-SVX samples
are shown in figure 2. The SVX sample and the non-SVX sample
contain $202 \pm 18$ and $193 \pm 26$ events respectively. The event
yields are extracted from an unbinned likelihood fit.
The total sample of $J/\psi K^0_S$ is $395 \pm 26$ events is used in this
analysis. The SVX subsample was used for a previous measurement of
$\sin 2\beta$\cite{oldcdf}.

\EPSFIGURE[ht]{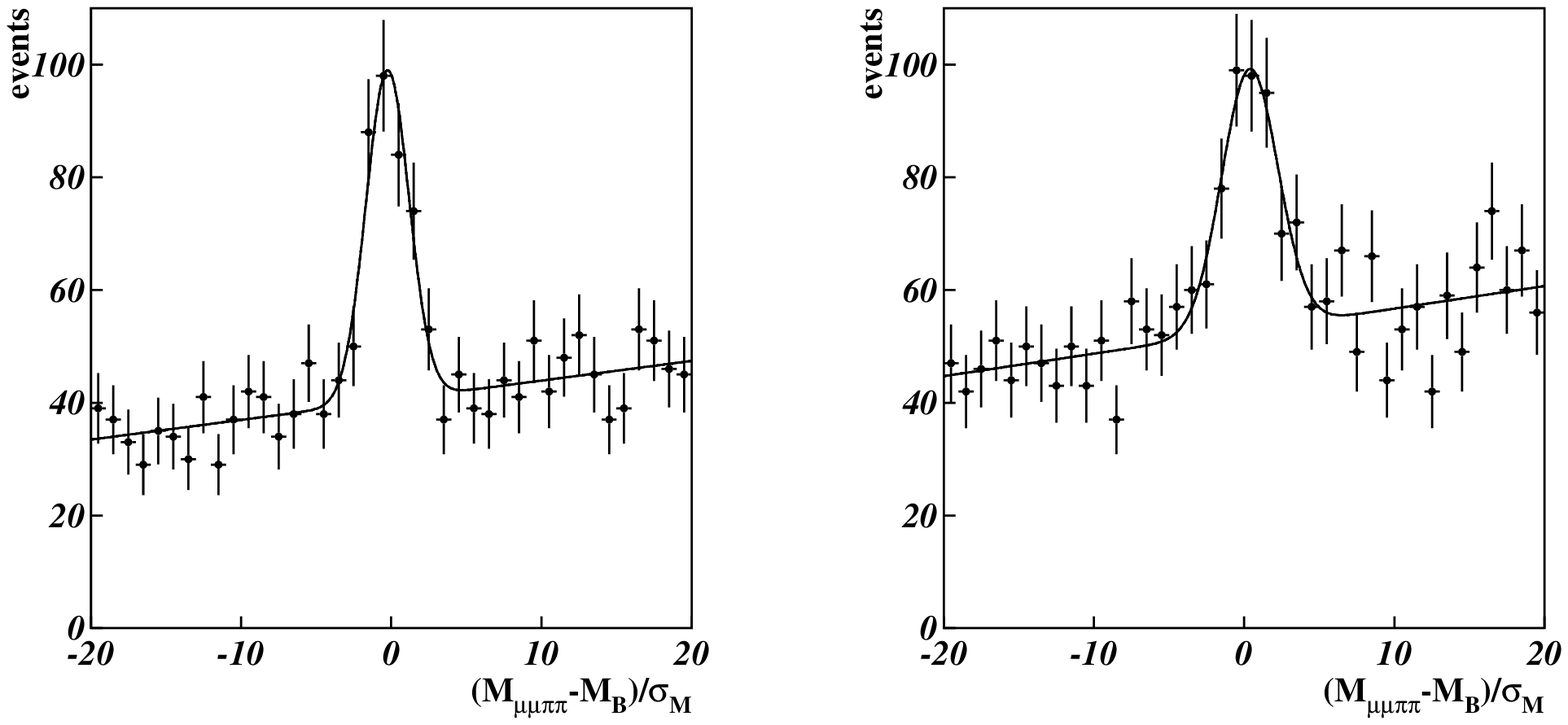,width=16cm}
{Left: Normalized mass for the $J/\psi K^0_S$  candidates where
both muons have high precision decay length measurements in the silicon
vertex
detector. Right: Normalized mass for the  $J/\psi K^0_S$
where either one or both muons are not measured in the silicon
vertex detector.}

\section{Flavor tagging}

In order to observe the CP asymmetry, $A_{CP}$, we must determine
the $b$ flavour at production by establishing whether the B meson
contains a $b$ or $\bar b$ quark.
Since the error on $\sin 2 \beta$ depends on $1/\sqrt{\epsilon D^2 N}$,
we can improve the statistical reach by using several taggers.
CDF has studied three tagging algorithms to measure $\Delta m_d$ in
$B^0-\bar B^0$ oscillations. CDF measures  $\Delta m_d=0.495 \pm 0.026 \pm
0.025$
ps$^{-1}$, which agrees well with the world average\cite{PDG}.
Two are opposite side tag algorithms and one is a same side tag algorithm.
The
dilution parameters for all tagging algorithms are measured
on calibration samples.
At the Tevatron the strong interaction creates $b \bar b$ pairs
at sufficiently high energy that the B mesons are largely
uncorrelated. For example, the
b quark could hadronize as a $\bar B^0$ while the
$\bar b$ could hadronize as a $ B^+$, $B^0$, or $B^0_s$ meson. Therefore
we can use a sample of $998 \pm 51$ $B^{\pm} \rightarrow J/\psi K^{\pm} $
decays to measure the tagging dilutions for the opposite side algorithms.
The performance of the same side tagging methods is evaluated by tagging
$B \rightarrow \nu \ell D^{(*)}$ decays and by measuring
the time dependence of $B^0 \bar B^0$ oscillations in this high statistics
sample ($\approx$ 6,000 events) and in a lower statistics sample of
$B \rightarrow J/\psi K^{*0}$ ($\approx$ 450 events). In the mixing case,
the measured asymmetry $A_{mix}$ is given by:
\begin{eqnarray*}
A_{mix}(t) & = &
\frac{N_{unmixed}(t)-N_{mixed}(t)}{N_{unmixed}(t)+N_{mixed}(t)} \\
& = & D cos (\Delta m_d t)
\end{eqnarray*}
where $N_{unmixed}(t)$ and $N_{mixed}(t)$ are the number of candidates
with same or opposite $b$ flavor. CDF has used the meaasurements of
$B^0-\bar B^0$
mixing to determine $D$ for the three different tagging methods used in
this analysis.

The same side tagging method or SST \cite{sst} relies on the correlation
between the $B$ flavor and the charge of a nearby particle. Such a
correlation can arise from the fragmentation processes which form a B
meson from a $\bar b$ quark as illustrated
in figure 3 and from
the decay of an excited B meson state ($B^{**}$).
\EPSFIGURE[ht]{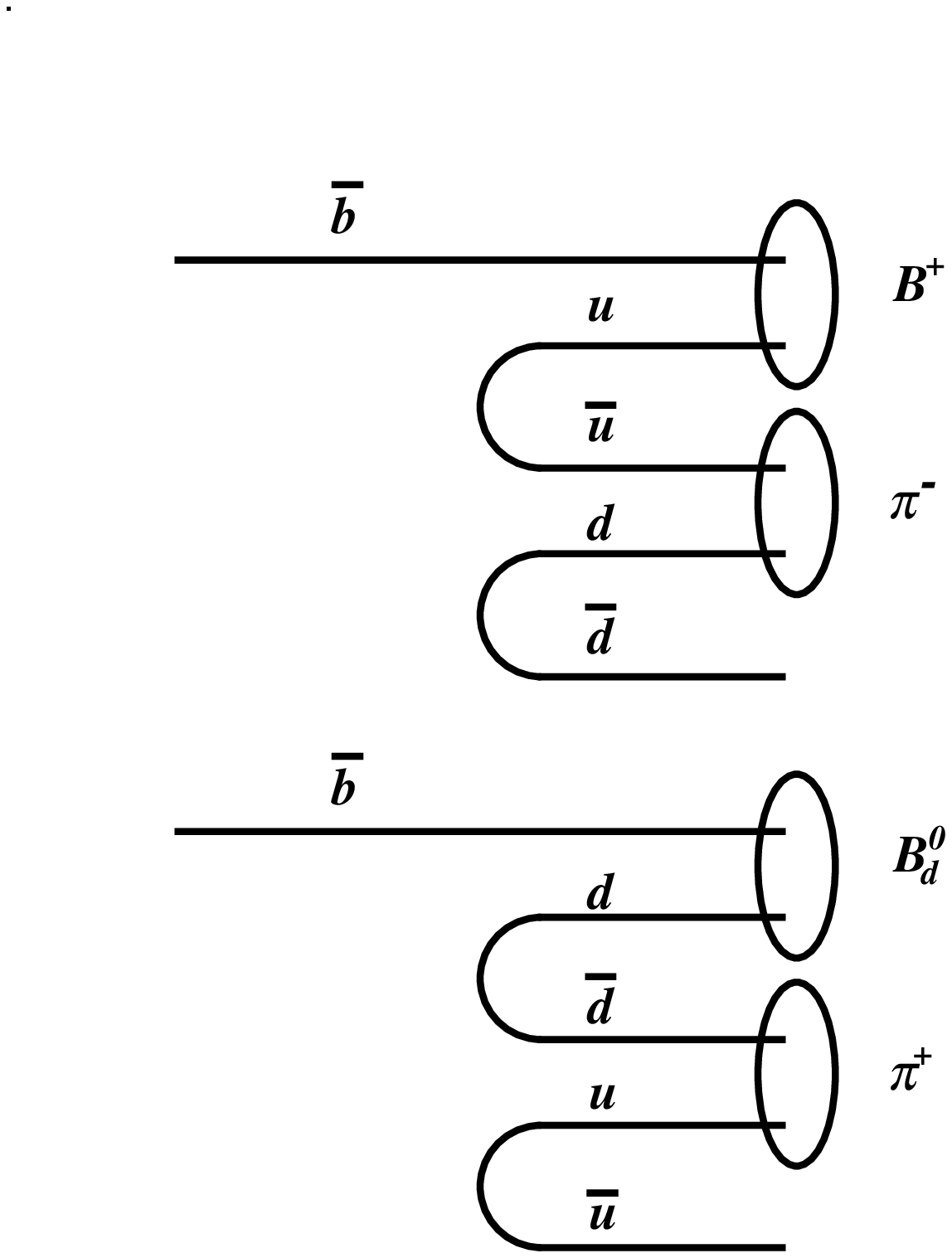,width=8cm}
{The same side flavour tag method is based on the correlation
between the $b$ flavor and the charge of particles produced
in the $b$ quark fragmentation.}
In the fragmentation a $\bar b$ quark forming a $ B^0$ can combine
with a $d$  in the hadronization leaving a $\bar d$ which can form a $\pi^+$
with a $u$ quark from the sea. The excited B state
will decay $B^{**+} \rightarrow B^{(*)0}\pi^+$.
Therefore in both cases a $B^0$ ($\bar B^0$) meson is associated with a
positive (negative) particle respectively. The effectiveness of this method
has been demonstrated by tagging $B \rightarrow \nu \ell D^{(*)}$ decays
and observing the time
dependence of the $B^0-\bar B^0$ oscillations.
The SST method selects a single charged particle as a flavor tag from those
within an $\eta-\phi$ cone of half-angle 0.7 around the B direction.
The pseudorapidity $\eta$ is defined as $\eta=-ln[\tan(\theta/2)]$;
$\theta$ is the polar angle relative to the proton beam direction,
and $\phi$ is the azimuthal angle around the beam line.
The tag must have $P_T$(tag)$>400$ MeV/c. It
must be reconstructed in the SVX and
have impact parameter within 3 $\sigma_d$ of the primary vertex,
where  $\sigma_d$ is the error on the impact parameter.
If there is more than one track candidate, we select as the flavor
tag the one with
the smallest $P^{rel}_T$, where $P^{rel}_T$ is the component of the
the track momentum with respect to the momentum of the combined
B+track system. The dilution of the SST sample
was measured, using the SVX sample, to be $D=(16.6 \pm 2.2 )$ \%.
This dilution was obtained by measuring $D$ in the large
$B\rightarrow \ell D^{(*)}X$ data sample used in the mixing analysis.
A Monte Carlo simulation was used to  extrapolate from the
momentum range of B mesons in semileptonic decays, which have an
average $P_T$ of 21 GeV/c, to the lower B momentum of the
$B\rightarrow J/\psi K^0_S$ sample with an average $P_T$ of 12 GeV/c.
When using the non-SVX sample, the SST tagging algorithm is modified
by dropping
the SVX information for all candidate tagging tracks and therefore
increasing the geometric acceptance. A dilution scale factor $f_D$
is defined by $D_{non-SVX}=f_D D_{SVX}$. The scale factor
is measured in the
$B^{\pm} \rightarrow J/\psi K^{\pm}$ sample to be $1.05 \pm 0.17$,
yielding a dilution in the non-SVX sample $D=(17.4 \pm 3.6)$ \%.

The opposite side tagging algorithms identify the flavor of the
{\it opposite}
B in the event at the time of production. CDF employs two opposite side
tagging methods: soft lepton tag (SLT) and jet-charge tag (JETQ).

The soft lepton tag \cite{slt} associates the charge of the lepton
($e$ or $\mu$) from semileptonic decays with the flavor of the parent B
meson as $b\rightarrow \ell^-$. Since we are tagging the opposite
B meson, its flavor is anti-correlated with the flavor of the B-meson that
decays to $J/\psi K^0_S$. Hence a $\ell^-$($\ell^+$) tags a $B^0$
($\bar B^0$) decaying as $B\rightarrow J/\psi K^0_S$.
A soft muon tag is defined as an identified muon with $P_T>2$ GeV/c; the
muon selection is similar to that for $J/\psi \rightarrow \mu^+\mu^-$.
A soft electron tag is defined as a charged track reconstructed in the
Central Tracking Chamber (CTC) with $P_T>1$ GeV/c and
extrapolated into the electromagnetic calorimeter.
The position information
from proportional chambers in the central calorimeter is required to match
the CTC track, and the shower profile and pulse height must be consistent with an
electron. The electron candidate's CTC track must have a $dE/dx$ deposition
consistent with an electron. A dilution of the SLT
tagging method is measured by applying the SLT algorithm to
the $B^{\pm} \rightarrow J/\psi K^{\pm}$ data sample.
We find $D=(62.5 \pm 14.6)$\%.

The other opposite side method, "Jet charge" or JETQ, tags the
b flavor by measuring the average charge of the opposite
side jet. If a soft lepton is not found we calculate $Q_{jet}$ as:
$$
Q_{jet}=\frac{\sum_i q_i P_{T_i}(2-(T_p)_i)}{\sum_i P_{T_i}(2-(T_p)_i)}
$$
where $q_i$ and $P_{T_i}$ are the charge and transverse momentum
of track $i$ in the jet. The quantity $T_P$ is the probability that
track $i$ originated from the $p\bar p$ interaction point. Tracks displaced
from the primary vertex are characterized
by $T_P \approx 0$. Therefore, displaced tracks from $B$ decays
will have a larger weight in the sum than tracks from the
primary interaction. For $b$ quark jets, the jet charge
has on average the same sign as the $b$ quark charge.
The algorithm was optimized by maximizing $\epsilon D^2$ on a sample
of $B^{\pm} \rightarrow J/\psi K^{\pm}$ events generated by a Monte Carlo
program. The jet is formed by clustering charged tracks with $P_T>0.40$
GeV/c
around a seed track of $P_T>1.75$ GeV/c until the mass of the cluster
is approximately equal to the mass of the B meson. Tracks within
$\Delta R<0.7$ are excluded to avoid overlap with SST tags.
The $B^0$ meson decay products are explicitly excluded from the jet.
If several jet candidates are found, we select the cluster which is most
likely to
be a $b$ jet based on impact parameter and cluster transverse momentum.
A $B^0$ ($\bar B^0$) is selected by $Q_{jet}<-0.2 ~(>0.2)$.
If the jet charge $|Q_{jet}|<0.2$, then the jet is considered untagged.
The dilution $D=(23.5\pm 6.9)$ \% is found by applying the JETQ algorithm
to the $B^{\pm} \rightarrow J/\psi K^{\pm}$ data sample.

Each event can be tagged by two tagging algorithms, one same side
and one opposite side. We use the SLT if both JETQ and SLT tags
are present to avoid correlations between the two opposite side
tagging algorithms.  A soft lepton tag SLT
can still be used as the SST track. This introduces a small
overlap between the two tagging algorithms which effects only
three events and does not change the final result.

A positive (+ tag) is defined
as the tag of a  $\bar b$ quark ($B^0$ meson). A negative ($-$tag)
is defined as the tag of a $b$ quark ($\bar B^0$ meson).
A null (0 tag) corresponds to an event where the flavor of the
B meson was not identified.
The efficiency and dilution of the tagging methods must be generalized
to take into account possible detector asymmetries. For example,
the CDF tracking chamber has a 1\% bias due to the
tilted cell geometry that favors the
reconstruction  of positive charge tracks at low transverse momentum.
To allow for this asymmetry, we define
$\epsilon^+_R$( $\epsilon^+_W$) as the fraction of B meson of type
+ that will be tagged as + ($-$). The fraction $\epsilon^+_0$ is the
fraction
of B meson of type + that are not tagged.
We have six parameters that account for these asymmetries-
$\epsilon^+_R$, $\epsilon^+_W$, $\epsilon^+_0$, $\epsilon^-_R$,
$\epsilon^-_W$ and $\epsilon^-_0$, but only four are independent
since by definition $\epsilon^+_R+ \epsilon^+_W+ \epsilon^+_0=1$
and $\epsilon^-_R+ \epsilon^-_W +\epsilon^-_0=1$.

The performances of the individual tagging methods are comparable
as summarized in table 1.
\TABULAR[ht]{ccc}
{Tagging method & $\epsilon$ & Dilution \\
SST SVX & $35.5 \pm 3.7$ &  $16.6 \pm 2.2$ \\
SST non-SVX & $38.1 \pm 3.9$ &  $17.4 \pm 3.6$ \\
SLT all & $5.6 \pm 1.8$ &  $62.5 \pm 14.6$ \\
JETQ all & $40.2 \pm 3.9$ &  $23.5 \pm 6.9$
}
{Summary of the tagging algorithms performance. All numbers
are in percent. The efficiencies are obtained from
the $B\rightarrow J/\psi K^0_S$ sample.
The dilution parameters
are derived from the $B^{\pm} \rightarrow J/\psi K^{\pm}$ sample
}
The three tagging methods are combined to reduce the uncertainty
in the CP asymmetry.
Since the soft lepton tagging has low efficiency but
high dilution, the jet charge tagging information is dropped if there is
a lepton tag. This avoids correlations between SST and JETQ. Therefore  a
neutral $B$ is tagged by at most two methods.
We combine two tagging algorithms by defining the dilution
weighted tags ${\cal D}_i=q_iD_i$ where $q=1,-1,0$  for $B^0$, $\bar B^0$
and untagged events respectively.
The combined dilution is:
$$
{\cal D}_{q_{1}q_{2}}=\frac{{\cal D}_1+{\cal D}_2}{1+{\cal D}_1{\cal D}_2}
$$
and the combined efficiency is:
$$
\epsilon_{q_{1}q_{2}}=\epsilon_{q_1}\epsilon_{q_2} (1+{\cal D}_1{\cal D}_2)
$$
If the tags agree, the effective dilution is increased while if they
disagree, the effective dilution is decreased.  If two perfect
tagging algorithms give an opposite tagging charge $q$
( ${D}_1={D}_2$ and $|{\cal D}_1|=1$), then the effective
dilution and the combined efficiency must be zero since
by definition perfect tagging algorithms cannot disagree.
The combined tagging power is
$\epsilon D^2=(6.3 \pm 1.7)$ \%, and the efficiency for flavor tagging
a $J/\psi K^0_S$ with a least one tag is $\approx 80$ \%.

\section{The measurement of $\sin(2\beta)$}

An unbinned likelihood fit is used to determine
$\sin 2\beta$. The parameters in the fit
can be described as a vector with 65 components.
The value of $\sin(2\beta)$ is a free parameter of the
fit while the remaining 64 parameters
describe other features of the data.
The likelihood function is described in detail in
\cite{newcdf}. The likelihood function is the product
$\prod_i {P_i}$ where $i$ runs over all selected events
and $P_i$ is the probability distribution
that an event is signal or background for the
normalized mass, the lifetime and the flavor
tag ( $q_1, q_2, q_3$ ).
There is a separate set of parameters for the SVX sample
and the non-SVX sample but both are part of the
same fit. The $P_i$ assume that the events
are signal or background, either prompt or long lived,
and therefore the $P_i$ contains three components
$P_S$, $P_P$ and $P_L$.
The fit includes the effective dilution, the
normalized mass and the lifetime.
Each of the components is expressed
as the product of a time-function
($T_S$, $T_P$, $T_L$), a mass-function
($M_S$, $M_P$, $M_L$), and the tagging efficiency
($\epsilon_S$, $\epsilon_P$, $\epsilon_L$).
The time-function $T_S$ has a dependence on the $B$ lifetime
and the mixing parameter.
The $B^0$ lifetime $\tau$ and the mixing parameter $\Delta m_d$
are constrained to the world averages:
$\tau=(1.54 \pm 0.04)$ ps and
$\Delta m_d=(0.464 \pm 0.018) \hbar$ ps$^{-1}$.
The prompt $J/\psi$ background is represented by a Gaussian
which depends on the proper time and its uncertainty.
The long lived background function $T_L$ has positive and negative
exponentials
in time which represent positive
and negative long-lived background.
The positive long-lived background
is from real $B$ decays while the negative
component is due to non-Gaussian tails
in the lifetime resolution.
The signal mass function $M_S$ is a Gaussian while the
prompt and long lived mass functions
are linear in mass.
The tagging efficiency function for the signal is constrained
by the measurement of the tagging efficiencies and dilutions.
The prompt and long-lived background functions, $\epsilon_P$
and $\epsilon_L$, give the probability
of obtaining the observed combination of tags for prompt and long-lived
background events.

\EPSFIGURE[htb]{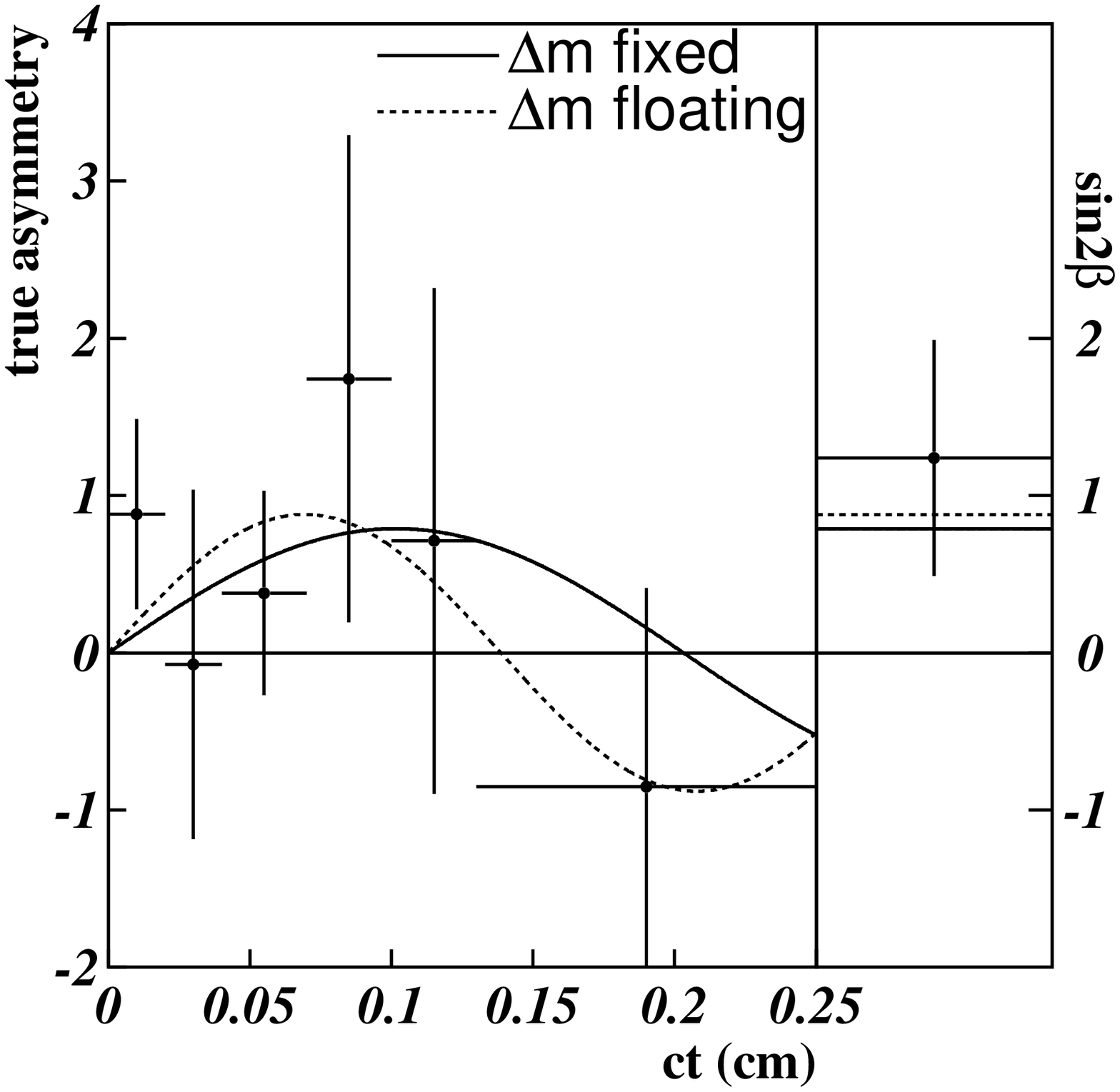,width=7cm}
{The true asymmetry as a function of time for $J/\psi K^0_S$ events.
The data points are side-band subtracted and have been combined
according to the effective dilutions for single and double tags.
The time integrated asymmetry for non-SVX events are shown on the
right.}

The fit yields $\sin(2\beta)=0.79^{+0.41}_{-0.44}$
$(stat. + syst.)$  The asymmetry is shown in figure 4 for the
SVX and non-SVX events separately. The asymmetry for the SVX events
which have good $ct$ resolution
is shown as a function of the proper lifetime.
The lifetime information for the non-SVX events is utilized in the
fit but the time-integrated asymmetry is shown in
figure 4 because the decay length information has low resolution.
The data shown in figure 4 have been side-band subtracted.
The curves displayed in the plot are the results of the full
maximum likelihood fit using all the data.
The uncertainty can be divided into statistical and
systematic terms:
$$
\sin (2 \beta)= 0.79 \pm 0.39 ({\rm stat.}) \pm 0.16 ({\rm syst.})
$$
The error is dominated by the statistical uncertainty.
The systematic term
is dominated by the uncertainty in the dilution parameters.

A scan of the likelihood function can be used  to determine whether
the CDF result supports $\sin(2\beta)> 0$.
Using the Feldman-Cousins frequentist approach
\cite{feld} we
determine $ 0.0< \sin 2 \beta < 1$ at 93 \% confidence level.
The Bayesian method assumes
a flat prior probability in $\sin (2\beta)$ and yields
$\sin (2 \beta) >0 $ at 95 \% confidence level.
Moreover if the true value of $\sin 2\beta $ is zero and
a Gaussian error uncertainty 0.44 is assumed, the probability
of measuring $\sin 2\beta >0.79$ is 3.6 \%.
This is the first compelling evidence for CP violation in $B$ meson
decays.

Several checks were performed.
We remove the constraint of $\Delta m_d$ to the world average
in the fit and measure
$\sin 2 \beta = 0.99 ^{+0.44}_{-0.41}$
and $\Delta m_d = 0.68 \pm 0.17 $ ps$^{-1}$.
A simplified time-integrated measurement of the asymmetry
yields $\sin 2 \beta= 0.71 \pm 0.63$.

The same tagging analysis and fitting procedure
was applied to a sample of $\approx 450$
$B^0\rightarrow J/\psi K^{*0}$ events.
The fit yields $\Delta m_d = 0.40 \pm 0.18 $ ps$^{-1}$
which is consistent with the world
average  $\Delta m_d = 0.464 \pm 0.018 $ ps$^{-1}$.

\section{Summary and future prospects}

The CDF measurement provides a first evidence for large
CP asymmetries in the $B^0$ system.
More precise measurements of $\sin 2 \beta$ will
soon be provided by the B-factories at SLAC and
KEK which have recently started data taking.
The CDF and D0 experiments at the Tevatron are expected
to resume data taking with the upgraded
Tevatron in March 2001. Run II should deliver
a factor of 20 more luminosity than run I. Moreover
the CDF and the D0 detectors will undergo
major upgrades\cite {d0talk}.
For CDF we project a data sample of 10,000 $ J/\psi K^0_S$
events from dimuon triggers yielding an
error on $\sin 2 \beta$ of about $\pm 0.08$.
CDF is considering adding a $J/\psi \rightarrow e^+ e^-$ trigger
which will increase the data sample by $\approx$ 50 \%.
A time of flight system which will
improve flavor tagging has also been added to the
run II upgrade. Moreover, the study of CP violation in $B_s$ decays
and the measurement of $\Delta m_s$ will be unique to hadron
colliders. Therefore, the Tevatron will continue to play a unique
role in the testing of the CKM matrix.

\section{Acknowledgments.}

I would like to thank the organizing committee for the
excellent meeting.

\end{document}